\documentclass[10pt, conference]{IEEEtran}
\usepackage{cite}
\usepackage{amsmath,amssymb,amsfonts}
\usepackage{algorithmic}
\usepackage{graphicx}
\usepackage{textcomp}
\usepackage{xcolor}
\usepackage{xspace}
\usepackage{url}
\usepackage{tcolorbox}
\usepackage{booktabs}
\usepackage{enumitem}
\usepackage{subcaption}
\usepackage{listings}
\usepackage{tikz}
\usepackage{array}

 \definecolor{pblue}{rgb}{0.13,0.13,1}
\definecolor{pgreen}{rgb}{0,0.5,0}
\definecolor{pred}{rgb}{0.9,0,0}
\definecolor{pgrey}{rgb}{0.46,0.45,0.48}
 \lstdefinelanguage{PrettyJava}
{ 
  language=Java,
  showspaces=false,
  showtabs=false,
  breaklines=true,
  showstringspaces=false,
  breakatwhitespace=true,
  commentstyle=\color{pgreen},
  keywordstyle=\color{pblue},
  basicstyle=\fontsize{6.5}{6.6} \ttfamily,
  moredelim=[is][\bfseries\textcolor{pgreen}]{/+}{+/},
  moredelim=[is][\bfseries\textcolor{pred}]{/-}{-/}
}

\def\BibTeX{{\rm B\kern-.05em{\sc i\kern-.025em b}\kern-.08em
    T\kern-.1667em\lower.7ex\hbox{E}\kern-.125emX}}

\newcolumntype{R}[1]{>{\raggedleft\let\newline\\\arraybackslash\hspace{0pt}}m{#1}}

\begin{document}

\newcommand{\vmm}{{\em Vulnerability-mimicking Mutants}\xspace}
\newcommand{\our}{{\em VMMS}\xspace}
\newcommand{\mubert}{{\em $\mu$BERT}\xspace}
\newcommand{\vulj}{{\em Vul4J}\xspace}

\title{Vulnerability Mimicking Mutants}

\author{
\IEEEauthorblockN{Aayush Garg}
\IEEEauthorblockA{\textit{University of Luxembourg} \\
\textit{Luxembourg} \\
aayush.garg@uni.lu}
\and
\IEEEauthorblockN{Renzo Degiovanni}
\IEEEauthorblockA{\textit{University of Luxembourg} \\
\textit{Luxembourg} \\
renzo.degiovanni@uni.lu}
\and
\IEEEauthorblockN{Mike Papadakis}
\IEEEauthorblockA{\textit{University of Luxembourg} \\
\textit{Luxembourg} \\
michail.papadakis@uni.lu}
\and
\IEEEauthorblockN{Yves Le Traon}
\IEEEauthorblockA{\textit{University of Luxembourg} \\
\textit{Luxembourg} \\
yves.letraon@uni.lu}
}

\maketitle
 
\begin{abstract}
With the increasing release of powerful language models trained on large code corpus (e.g. CodeBERT was trained on 6.4 million programs), a new family of mutation testing tools has arisen with the promise to generate more ``natural'' mutants in the sense that the mutated code aims at following the implicit rules and coding conventions typically produced by programmers. 
In this paper, we study to what extent the mutants produced by language models can semantically mimic the observable behavior of security-related vulnerabilities (a.k.a. \vmm), so that designing test cases that are failed by these mutants will help in tackling mimicked vulnerabilities. 
Since analyzing and running mutants is computationally expensive, it is important to prioritize those mutants that are more likely to be vulnerability mimicking prior to any analysis or test execution. 
Taking this into account, we introduce \our, a machine learning based approach that automatically extracts the features from mutants and predicts the ones that mimic vulnerabilities. We conducted our experiments on a dataset of 45 vulnerabilities and found that 16.6\% of the mutants fail one or more tests that are failed by 88.9\% of the respective vulnerabilities. More precisely, \emph{3.9\%} of the mutants from the entire mutant set are vulnerability-mimicking mutants that mimic \emph{55.6\%} of the vulnerabilities. Despite the scarcity, \our predicts vulnerability-mimicking mutants with 0.63 MCC, 0.80 Precision, and 0.51 Recall, demonstrating that the features of vulnerability-mimicking mutants can be automatically learned by machine learning models to statically predict these without the need of investing effort in defining such features. 
\end{abstract}



\section{Introduction}
\label{sec_introduction}
Research and practice with mutation testing have shown that it is one of the most powerful testing techniques~\cite{DBLP:journals/computer/DeMilloLS78, DBLP:books/daglib/0020331, 9677967, DBLP:journals/stvr/Reid99}. 
Apart from testing the software in general, mutation testing has been proven to be useful in supporting many software engineering activities which include improving test suite strength~\cite{DBLP:conf/icse/ChekamPTH17, DBLP:conf/icst/AmmannDO14}, selecting quality software specifications~\cite{DBLP:conf/sigsoft/TerragniJTP20, DBLP:conf/icse/MolinaPAF21, DBLP:conf/icse/MolinadA22}, among others. 
Though, its use in tackling software security issues has received little attention. 
A few works focused on model-based testing~\cite{DBLP:conf/tap/BuchlerOP11,DBLP:conf/models/MouelhiFBT08} and proposed security-specific mutation operators to inject potential security-specific leaks into models that can lead to test cases to find attack traces in internet protocol implementations. 
Other works proposed new security-specific mutation operators that aim to mimic common security bug patterns in Java~\cite{DBLP:conf/icst/LoiseDPPH17} and C~\cite{DBLP:conf/icst/NanavatiWHJK15}. 
These works empirically showed that traditional mutation operators are unlikely to exercise security-related aspects of the applications and thus, the proposed operators attempt to convert non-vulnerable code to vulnerable by mimicking common real-world security bugs. 
However, pattern-based approaches have two major limitations. 
On one hand, the design of security-specific mutation operators is not a trivial task since it requires manual analysis and comprehension of the vulnerability classes that cannot be easily expanded to the extensive set of realistic vulnerability types (e.g. they restrict to memory~\cite{DBLP:conf/icst/LoiseDPPH17} and web application~\cite{DBLP:conf/icst/NanavatiWHJK15} bugs). 
On the other hand, these mutation operators can alter the program semantics in ways that 
may not be convincing for developers as they may perceive them as unrealistic/uninteresting \cite{BellerWBSM0021}, thereby hindering the usability of the method.  

With the aim of producing more realistic and natural code, a new family of tools based on language models has recently arisen. 
Currently, language models are employed for code completion~\cite{10.1145/3324884.3416591}, test oracle generation~\cite{DBLP:conf/ast/TufanoDSS22}, program repair~\cite{DBLP:journals/tse/ChenKTPPM21}, among many other software engineering tasks. 
Particularly, language models are been used for mutant generation yielding to several mutation testing tools such as SemSeed~\cite{SemSeed} and DeepMutation~\cite{DeepMutation}. While these tools are subject to expensive training on datasets containing thousands of buggy code examples, there is an increasing interest in using pre-trained language models for mutant generation~\cite{9787824,DBLP:journals/corr/abs-2206-01335,DBLP:conf/icst/DegiovanniP22}, e.g. a mutation testing tool \mubert~\cite{DBLP:conf/icst/DegiovanniP22} uses CodeBERT~\cite{DBLP:conf/emnlp/FengGTDFGS0LJZ20} to generate mutants by masking and replacing tokens with CodeBERT predictions. 

Since pre-trained language models were trained on large code corpus (e.g. CodeBERT was trained on more than 6.4 million programs), their predictions are typically considered representative of the code produced by programmers. 
Hence, we wonder:

\noindent
\emph{Are mutation testing tools using pre-trained language models effective at producing mutants that semantically mimic  the behaviour of software vulnerabilities?     
}

A positive answer to this question can be a promising prospect for the use of these security-related mutants to form an initial step for defining security-conscious testing requirements. 
We believe that these requirements are particularly useful when building regression test suites for security intensive applications. 


The task of analyzing the mutants, and writing and executing tests, in general, is considered expensive. 
Despite a large number of mutants created, it is well known that many of them are of low utility, i.e., they do not contribute much to the testing process~\cite{DBLP:journals/infsof/JiaH09, DBLP:conf/apsec/KintisPM10, DBLP:conf/icst/AmmannDO14}. 
Due to this, several mutant selection techniques have been proposed to make mutation testing more cost-effective~\cite{Papadakis+2019,DBLP:journals/tosem/OffuttLRUZ96,DBLP:conf/kbse/ZhangGMK13,DBLP:conf/sigsoft/KurtzAODKG16,ChekamPBTS20}. 
Therefore, to make our approach useful in practice, we need to filter and select only specific mutants that resemble the behavior of security issues, especially vulnerabilities.

Taking this into account, we propose to enable security-conscious mutation testing by focusing on a minimal set of mutants that rather behave similarly to vulnerabilities a.k.a. \vmm. Such mutants are the ones that semantically mimic the observable behavior of vulnerabilities, i.e., a mutant is vulnerability-mimicking when it fails the same tests that are failed by the vulnerability that it mimics, proving its existence in the software a.k.a. \emph{PoV} (Proof of Vulnerability). 
Using \vmm as test requirements can guide testers to design test suites for tackling vulnerabilities similar to the mimicked ones. 

We conducted experiments on a dataset of 45 reproducible vulnerabilities, with severity ranging from high to medium, and found that for 40 out of 45 vulnerabilities,  (i.e., for 88.9\% vulnerabilities) there exists at least one mutant that fails one or more tests that are also failed by the respective vulnerabilities. More precisely, \emph{3.9\%} of the mutants from the entire mutant set are vulnerability-mimicking. Despite being few in quantity, these \vmm semantically mimic \emph{55.6\%} of the vulnerabilities, i.e., these mutants fail the ``same" tests that are failed by the respective vulnerabilities that they mimicked.

Since such mutants are very few among the large set of mutants generated, 
we propose \our\footnote{\emph{Vulnerability Mimicking Mutant Selector (VMMS)}}, a machine learning based approach that automatically learns the features of \vmm to identify these mutants \emph{statically}. \our is very accurate in predicting \vmm with \emph{0.63 MCC, 0.80 Precision, and 0.51 Recall}. This demonstrates that the features of \vmm can be automatically learned by machine learning models to statically predict these without the need of investing effort in manually defining any related features. 
We believe that \vmm can help in building regression test suites for security intensive applications, and can be particularly useful in evaluating and comparing fuzzing or other security testing tools. 
In summary, our paper makes the following contributions:

\begin{enumerate}
    \item We show that mutation testing tools based on language models can generate mutants that mimic real software vulnerabilities. 3.6\% of the mutants semantically mimic 25 out of 45 studied vulnerabilities. 
    \item We also show that for most of the vulnerabilities (40 out of 45) there exists at least one mutant that fails the one test finding the vulnerability (although not mimicking it).
    \item We propose \our, a machine-learning based approach for identifying \vmm. Our results show that \our is very accurate in its predictions as it obtains 0.63 MCC, 0.80 Precision, and 0.51 Recall.
\end{enumerate}


\section{Background}
\label{sec_background}
\subsection{Mutation Testing}
Mutation testing is a popular fault-based testing technique~\cite{DBLP:journals/computer/DeMilloLS78, DBLP:books/daglib/0020331}. 
It works by introducing slight syntactic modifications to the original program, a.k.a., \emph{mutants}. These mutants are artificially seeded faults that aim at simulating bugs present in the software. The tester designs test cases in order to \emph{kill} these mutants, i.e., to distinguish the observable behavior between a mutant and the original program. Thus, selecting specific mutants enables testing specific structures of a given language that the testing process seeks~\cite{9677967}. Due to this flexibility, Mutation Testing is used to guide test generation~\cite{DBLP:conf/issre/PapadakisM10}, to perform test assessment~\cite{PapadakisHHJT16}, to uncover subtle faults~\cite{DBLP:conf/icse/ChekamPTH17}, and to perform strong assertion inference~\cite{DBLP:conf/icse/MolinadA22}.

\subsection{Vulnerabilities}
Common Vulnerability Exposures (CVE)~\cite{CVE:Glossary} defines a security vulnerability as ``\emph{a flaw in a software, firmware, hardware, or service component resulting from a weakness that can be exploited, causing a negative impact to the confidentiality, integrity, or availability of an impacted component or components.}''. The inadvertence of a developer or insufficient knowledge of defensive programming usually causes these weaknesses. Vulnerabilities are usually reported in publicly available databases to promote their disclosure and fix. One such example is National Vulnerability Database, aka NVD~\cite{NVD}. NVD is the U.S. government repository of standards based vulnerability management data. All vulnerabilities in the NVD have been assigned a CVE (Common Vulnerabilities and Exposures) identifier. The Common Vulnerabilities and Exposures (CVE) Program’s primary purpose is to uniquely identify vulnerabilities and to associate specific versions of codebases (e.g., software and shared libraries) to those vulnerabilities. The use of CVEs ensures that two or more parties can confidently refer to a CVE identifier (ID) when discussing or sharing information about a unique vulnerability. 

\subsection{\vmm}
\label{sec_vmm}
The issues related to security, especially vulnerabilities have received less attention in the mutation testing literature. As a result, despite its flexibility, mutation testing has not been used as the first line of defense against vulnerabilities. Also, there is no clear definition of \vmm, (i.e., mutants that mimic the vulnerability behavior) to focus on, in order to perform mutation testing to guarantee the software under analysis is vulnerability-free. Therefore, for the purpose of this study, we use the following definition: 

\noindent
\emph{A mutant is vulnerability-mimicking if it fails exactly the same tests that are failed by the vulnerability it mimics, hence having the same observable behavior as the vulnerability.}

Since a mutant is a slight syntactic modification to the original program, a large number of mutants are generated during mutation testing which requires analysis and execution with the related test suites. This introduces a problem of identifying \vmm among a huge pile of mutants. In our dataset, \vmm are \emph{3.9\%} of the entire lot. To deal with the problem of identification of \vmm, we introduce \our, a deep learning based approach that predicts \vmm without requiring any dynamic analysis.

\begin{table*}[t!]
\caption{The table records the Vulnerability dataset details that include CVE ID, CWE ID and description, Severity level (that ranges from 0 to 10), number of Files and Methods that were modified during the vulnerability fix, and number of Tests that are failed by the vulnerability a.k.a. Proof of Vulnerability (PoV).}
\begin{center}
\begin{tabular}{l|l|l|c|c|c|c}
\toprule
\multicolumn{1}{c|}{\textbf{CVE}} & \multicolumn{1}{c|}{\textbf{CWE}} & \multicolumn{1}{c|}{\textbf{CWE description}} & \multicolumn{1}{c|}{\textbf{Severity}} & \multicolumn{1}{c|}{\textbf{$\#$ Files}} & \multicolumn{1}{c|}{\textbf{$\#$Methods}} & \multicolumn{1}{c}{\textbf{Failed Tests}} \\
\multicolumn{1}{c|}{\textbf{(Vulnerability)}} &  & \multicolumn{1}{c|}{\textbf{(Common Weakness Enumeration)}} & \multicolumn{1}{c|}{\textbf{(0 - 10)}} & \multicolumn{1}{c|}{\textbf{modified}} & \multicolumn{1}{c|}{\textbf{modified}} & \multicolumn{1}{c}{\textbf{(PoV)}} \\ \midrule
CVE-2017-18349 & CWE-20 & Improper Input Validation & 9.8 & 1 & 1 & 1 \\
CVE-2013-2186 & CWE-20 & Improper Input Validation & 7.5 & 1 & 1 & 2 \\
CVE-2014-0050 & CWE-264 & Permissions, Privileges, and Access Controls & 7.5 & 2 & 5 & 1 \\
CVE-2018-17201 & CWE-835 & Loop with Unreachable Exit Condition ('Infinite Loop') & 7.5 & 1 & 1 & 1 \\
CVE-2015-5253 & CWE-264 & Permissions, Privileges, and Access Controls & 4.0 & 1 & 1 & 1 \\
HTTPCLIENT-1803 & NA & NA & NA & 1 & 1 & 1 \\
PDFBOX-3341 & NA & NA & NA & 1 & 1 & 1 \\
CVE-2017-5662 & CWE-611 & Improper Restriction of XML External Entity Reference & 7.3 & 1 & 2 & 1 \\
CVE-2018-11797 & NA & NA & 5.5 & 1 & 1 & 1 \\
CVE-2016-6802 & CWE-284 & Improper Access Control & 7.5 & 1 & 1 & 3 \\
CVE-2016-6798 & CWE-611 & Improper Restriction of XML External Entity Reference & 9.8 & 1 & 2 & 1 \\
CVE-2017-15717 & CWE-79 & Improper Neutralization of Input During Web & 6.1 & 1 & 2 & 2 \\
 &  & Page Generation ('Cross-site Scripting') &  &  &  &  \\
CVE-2016-4465 & CWE-20 & Improper Input Validation & 5.3 & 1 & 1 & 1 \\
CVE-2014-0116 & CWE-264 & Permissions, Privileges, and Access Controls & 5.8 & 1 & 4 & 1 \\
CVE-2016-8738 & CWE-20 & Improper Input Validation & 5.8 & 1 & 1 & 2 \\
CVE-2016-4436 & NA & NA & 9.8 & 1 & 2 & 1 \\
CVE-2016-2162 & CWE-79 & Improper Neutralization of Input During Web & 6.1 & 1 & 2 & 1 \\
 &  & Page Generation ('Cross-site Scripting') &  &  &  &  \\
CVE-2018-8017 & CWE-835 & Loop with Unreachable Exit Condition ('Infinite Loop') & 5.5 & 1 & 2 & 1 \\
CVE-2014-4172 & CWE-74 & Improper Neutralization of Special Elements in Output & 9.8 & 2 & 2 & 1 \\
 &  & Used by a Downstream Component ('Injection') &  &  &  &  \\
CVE-2019-3775 & CWE-287 & Improper Authentication & 6.5 & 1 & 1 & 1 \\
CVE-2018-1002200 & CWE-22 & Improper Limitation of a Pathname to a Restricted & 5.5 & 1 & 1 & 1 \\
 &  & Directory ('Path Traversal') &  &  &  &  \\
CVE-2017-1000487 & CWE-78 & Improper Neutralization of Special Elements used & 9.8 & 3 & 17 & 12 \\
 &  & in an OS Command ('OS Command Injection') &  &  &  &  \\
CVE-2018-20227 & CWE-22 & Improper Limitation of a Pathname to a Restricted & 7.5 & 1 & 5 & 1 \\
 &  & Directory ('Path Traversal') &  &  &  &  \\
CVE-2013-5960 & CWE-310 & Cryptographic Issues & 5.8 & 1 & 2 & 15 \\
CVE-2018-1000854 & CWE-74 & Improper Neutralization of Special Elements in Output & 9.8 & 1 & 2 & 1 \\
 &  & Used by a Downstream Component ('Injection') &  &  &  &  \\
CVE-2016-3720 & NA & NA & 9.8 & 1 & 1 & 1 \\
CVE-2016-7051 & CWE-611 & Improper Restriction of XML External Entity Reference & 8.6 & 1 & 1 & 1 \\
CVE-2018-1000531 & CWE-20 & Improper Input Validation & 7.5 & 1 & 1 & 1 \\
CVE-2018-1000125 & CWE-20 & Improper Input Validation & 9.8 & 1 & 4 & 1 \\
APACHE-COMMONS-001 & NA & NA & NA & 1 & 1 & 1 \\
CVE-2013-4378 & CWE-79 & Improper Neutralization of Input During Web & 4.3 & 1 & 1 & 1 \\
 &  & Page Generation ('Cross-site Scripting') &  &  &  &  \\
CVE-2018-1000865 & CWE-269 & Improper Privilege Management & 8.8 & 1 & 3 & 1 \\
CVE-2018-1000089 & CWE-532 & Insertion of Sensitive Information into Log File & 7.4 & 1 & 2 & 1 \\
CVE-2015-6748 & CWE-79 & Improper Neutralization of Input During Web & 6.1 & 1 & 1 & 1 \\
 &  & Page Generation ('Cross-site Scripting') &  &  &  &  \\
CVE-2016-10006 & CWE-79 & Improper Neutralization of Input During Web & 6.1 & 1 & 1 & 1 \\
 &  & Page Generation ('Cross-site Scripting') &  &  &  &  \\
CVE-2018-1000615 & NA & NA & 7.5 & 1 & 1 & 1 \\
CVE-2017-8046 & CWE-20 & Improper Input Validation & 9.8 & 2 & 5 & 1 \\
CVE-2018-11771 & CWE-835 & Loop with Unreachable Exit Condition ('Infinite Loop') & 5.5 & 1 & 1 & 2 \\
CVE-2018-15756 & NA & NA & 7.5 & 1 & 5 & 2 \\
CVE-2018-1000850 & CWE-22 & Improper Limitation of a Pathname to a Restricted & 7.5 & 1 & 2 & 3 \\
 &  & Directory ('Path Traversal') &  &  &  &  \\
CVE-2017-1000207 & CWE-502 & Deserialization of Untrusted Data & 8.8 & 1 & 3 & 1 \\
CVE-2019-10173 & CWE-502 & Deserialization of Untrusted Data & 9.8 & 1 & 7 & 1 \\
CVE-2019-12402 & CWE-835 & Loop with Unreachable Exit Condition ('Infinite Loop') & 7.5 & 1 & 1 & 1 \\
CVE-2020-1953 & NA & NA & 10.0 & 1 & 7 & 2 \\
\bottomrule
\end{tabular}
\label{tab_data}
\end{center}
\end{table*}

\subsection{\vulj}
\label{sec_vulj}
There exist several vulnerability datasets for many programming languages~\cite{DBLP:conf/promise/BhandariNM21, DBLP:conf/msr/FanL0N20, DBLP:journals/ese/GargDJCPT22}. However, they do not contain bug-witnessing test cases to reproduce vulnerabilities, i.e., Proof of Vulnerability (PoV). Such test cases are essential for this study in order to determine whether generated mutants are \vmm, as explained in the section above. In general, it is hard to reproduce exploitation (i.e., PoV) for vulnerabilities. \vulj~\cite{DBLP:conf/msr/BuiSF22} is a dataset of real vulnerabilities, with the corresponding fixes and the PoV test cases, that we utilized for this study. Although, due to a few test cases failing even after applying the provided vulnerability-fixes, we had to exclude a few vulnerabilities. In total, we conducted this study on 45 vulnerabilities. In table~\ref{tab_data}, we mention the details of considered vulnerabilities that include CVE ID, CWE ID and description, Severity level (that ranges from 0 to 10, provided by National Vulnerability Database~\cite{NVD}), number of Files and Methods that were modified during the vulnerability fix, and number of Tests that are failed by the vulnerability a.k.a. Proof of Vulnerability (PoV).

\subsection{\mubert}
\label{sec_mubert}
\mubert~\cite{DBLP:conf/icst/DegiovanniP22} is a mutation testing tool that uses a pre-trained language model \emph{CodeBERT} to generate mutants by masking and replacing tokens. \mubert takes a Java class and extracts the expressions to mutate. It then masks the token of interest, e.g. a variable name, and invokes CodeBERT to complete the masked sequence (i.e., to predict the missing token). 
This approach has been proven efficient in increasing the fault detection of test suites~\cite{DBLP:conf/icst/DegiovanniP22} and improving the accuracy of learning-based bug-detectors~\cite{9787824} and thus, we consider it as a representative of pre-trained language-model-based techniques. 
For instance, consider the sequence \texttt{int total = out.length;} taken form Figure~\ref{fig_example_1_a}, \mubert mutates the object field access expression \texttt{length} by feeding CodeBERT with the masked sequence \texttt{int total = out.<mask>;}. CodeBERT predicts the 5 most likely tokens to replace the masked one, e.g.,  it predicts \texttt{total}, \texttt{length}, \texttt{size},  \texttt{count} and \texttt{value} for the given masked sequence. \mubert takes these predictions and generates mutants by replacing the masked token with the predicted ones (per masked token creates five mutants). \mubert discards non-compilable mutants and those syntactically the same as the original program (cases in which CodeBERT predicts the original masked token).


\begin{figure*}[t!]
\begin{subfigure}{0.32\textwidth}
    \begin{lstlisting}[language=PrettyJava]
private static void decompress
 (final InputStream in, final byte[] out)
 throws IOException {
  int position = 0;
  final int total = out.length;
  while (position < total) {
   final int n = in.read();


   
   if (n > 128) {
    final int value = in.read();
    for (int i = 0; i < (n & 0x7f); i++) {
     out[position++] = (byte) value; }
   } else {
    for (int i = 0; i < n; i++) {
    out[position++] = (byte) in.read();}
   }
  }
 }
\end{lstlisting}
    \caption{Vulnerable Code (CVE-2018-17201)}
    \label{fig_example_1_a}
  \end{subfigure}
\hfill
\begin{subfigure}{0.32\textwidth}
    \begin{lstlisting}[language=PrettyJava]
private static void decompress
 (final InputStream in, final byte[] out)
 throws IOException {
  int position = 0;
  final int total = out.length;
  while (position < total) {
   final int n = in.read();
   /+if (n < 0) {
    throw new ImageReadException("Error decompressing RGBE file"); }+/
   if (n > 128) {
    final int value = in.read();
    for (int i = 0; i < (n & 0x7f); i++) {
     out[position++] = (byte) value; }
   } else {
    for (int i = 0; i < n; i++) {
    out[position++] = (byte) in.read();}
   }
  }
 }
\end{lstlisting}
    \caption{Fixed Code}
    \label{fig_example_1_b}
  \end{subfigure}
\hfill
\begin{subfigure}{0.32\textwidth}
\begin{lstlisting}[language=PrettyJava]
private static void decompress
 (final InputStream in, final byte[] out)
 throws IOException {
  int position = 0;
  final int total = out.length;
  while (position < total) {
   final int n = in.read();
   if (n /-==-/ 0) { // `<' modified to `=='
    throw new ImageReadException("Error decompressing RGBE file"); }
   if (n > 128) {
    final int value = in.read();
    for (int i = 0; i < (n & 0x7f); i++) {
     out[position++] = (byte) value; }
   } else {
    for (int i = 0; i < n; i++) {
    out[position++] = (byte) in.read();}
   }
  }
 }
\end{lstlisting}
    \caption{Vulnerability-mimicking Mutant}
    \label{fig_example_1_c}
  \end{subfigure}
\caption{Vulnerability CVE-2018-17201 (Fig. \ref{fig_example_1_a}) that allows ``Infinite Loop" making code hang, which further enables Denial-of-Service (DoS) attack is fixed with the conditional exception using ``\texttt{if}" expression (Fig. \ref{fig_example_1_b}). Vulnerability-mimicking Mutant (Fig. \ref{fig_example_1_c}) modifies the ``\texttt{if}" condition that nullifies the fix and re-introduces the vulnerability.}
\label{fig_example_1}
\end{figure*}

\begin{figure*}[t!]
\begin{subfigure}{0.32\textwidth}
    \begin{lstlisting}[language=PrettyJava]
void addPathParam(String name, String value, boolean encoded) {
 if (relativeUrl == null) {
  throw new AssertionError(); }




  
 /-relativeUrl = relativeUrl.replace("{" + name + "}", canonicalizeForPath(value, encoded));-/

 

 
}
\end{lstlisting}
    \caption{Vulnerable Code (CVE-2018-1000850)}
    \label{fig_example_2_a}
  \end{subfigure}
\hfill
\begin{subfigure}{0.32\textwidth}
    \begin{lstlisting}[language=PrettyJava]
void addPathParam(String name, String value, boolean encoded) {
 if (relativeUrl == null) {
  throw new AssertionError(); }
 /+String replacement = canonicalizeForPath(value, encoded);
 String newRelativeUrl = relativeUrl.replace("{" + name + "}", replacement);
 if (PATH_TRAVERSAL
  .matcher(newRelativeUrl)
  .matches()) {
   throw new IllegalArgumentException(
   "@Path parameters shouldn't perform path traversal ('.' or '..'): " + value); }
 relativeUrl = newRelativeUrl;+/
}
\end{lstlisting}
    \caption{Fixed Code}
    \label{fig_example_2_b}
  \end{subfigure}
\hfill
\begin{subfigure}{0.32\textwidth}
\begin{lstlisting}[language=PrettyJava]
void addPathParam(String name, String value, boolean encoded) {
 if (relativeUrl == null) {
  throw new AssertionError(); }
 String replacement = canonicalizeForPath(value, encoded);
 String newRelativeUrl = relativeUrl.replace("{" + name + "}", replacement);
 if (PATH_TRAVERSAL
  .matcher(/-name-/)//passed argument changed
  .matches()) {
   throw new IllegalArgumentException(
   "@Path parameters shouldn't perform path traversal ('.' or '..'): " + value); }
 relativeUrl = newRelativeUrl;
}
\end{lstlisting}
    \caption{Vulnerability-mimicking Mutant}
    \label{fig_example_2_c}
  \end{subfigure}
\caption{Vulnerability CVE-2018-1000850 that allows ``Path Traversal", which further enables access to a Restricted Directory (Fig. \ref{fig_example_2_a}) is fixed with the conditional exception in case \texttt{`.'} or \texttt{`..'} appears in the ``\texttt{newRelativeUrl}" (Fig. \ref{fig_example_2_b}). Vulnerability-mimicking Mutant (Fig. \ref{fig_example_2_c}) in which the passed argument is changed from ``\texttt{newRelativeUrl}" to ``\texttt{name}" nullifies the fix and re-introduces the vulnerability.}
\label{fig_example_2}
\end{figure*}

\section{Motivating Examples}
\label{sec_motivation}
Figures \ref{fig_example_1} and \ref{fig_example_2} show motivating examples of how generated mutants can mimic the behavior of vulnerabilities. Fig.~\ref{fig_example_1} demonstrates the example of high severity (7.5) vulnerability CVE-2018-17201\cite{CVE-2018-17201} that allows ``Infinite Loop", a.k.a., a loop with unreachable exit condition when parsing input files. This makes the code hang which allows an attacker to perform a Denial-of-Service (DoS) attack. The vulnerable code (Fig. \ref{fig_example_1_a}) is fixed with the use of an ``\texttt{if}" expression (Fig. \ref{fig_example_1_b}) to throw an exception and moves out of the loop in case of such an event. Fig. \ref{fig_example_1_c} shows one of \vmm in which the ``\texttt{if}" condition is modified, i.e., the binary operator ``$\texttt{<}$" is modified to ``\texttt{==}". This modification makes the ``\texttt{if}" condition never executed, nullifying the fix, and behaving exactly the same as the vulnerable code. 

Fig.~\ref{fig_example_2} demonstrates the example of another high severity vulnerability CVE-2018-1000850\cite{CVE-2018-1000850} that allows ``Directory Traversal" that can result in an attacker manipulating the URL to add or delete resources otherwise unavailable to him/her. The vulnerable code (Fig. \ref{fig_example_2_a}) is fixed with the use of an ``\texttt{if}" expression (Fig. \ref{fig_example_2_b}) to throw an exception in case \texttt{`.'} or \texttt{`..'} appears in the ``\texttt{newRelativeUrl}" (Fig. \ref{fig_example_2_b}). Fig. \ref{fig_example_2_c} shows one of \vmm in which the passed argument is changed from ``\texttt{newRelativeUrl}" to ``\texttt{name}" which changes the matching criteria, hence nullifying the fix, and introducing same vulnerability behaviour.

\begin{figure*}[htp]
\centerline{\includegraphics[width=0.98\textwidth]{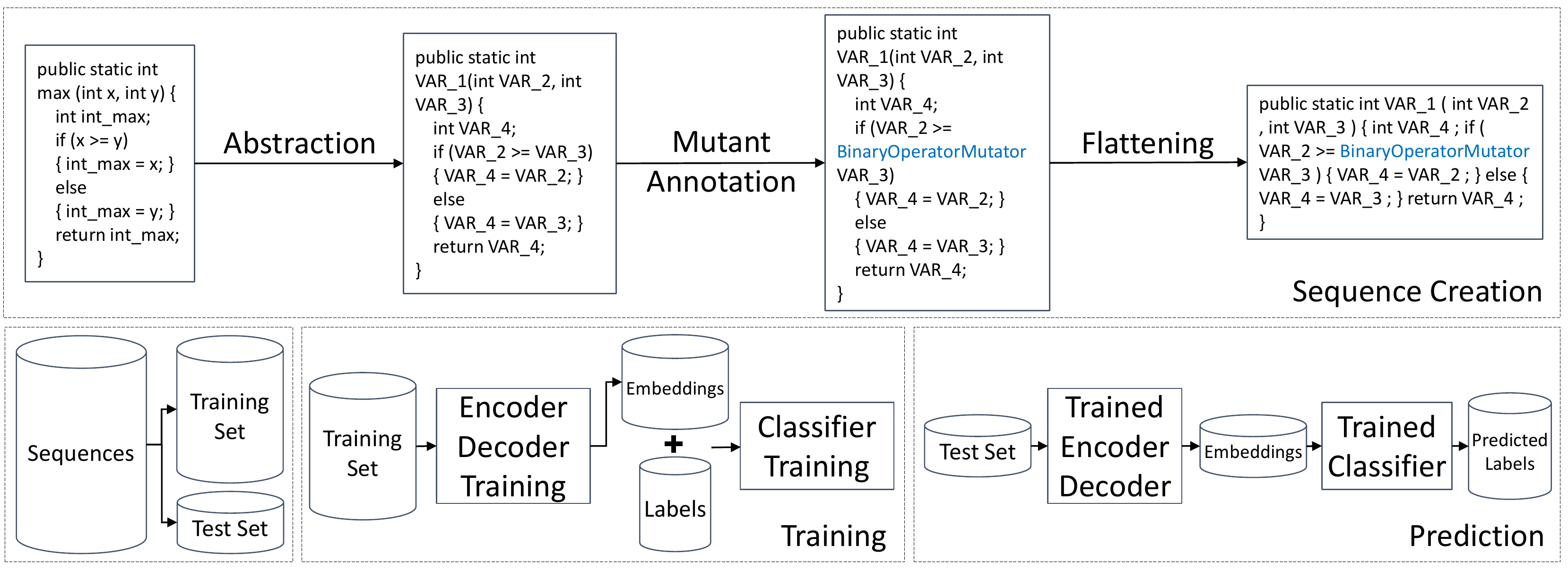}}
\vspace{-0.4em}
\caption{Overview of \our: Source code is abstracted and annotated to represent a mutant which is further flattened to create a single-space-separated sequence of tokens. An encoder-decoder model is trained on sequences to generate mutant embeddings. A classifier is trained on these embeddings and their corresponding labels (whether or not the mutants are \vmm). The trained classifier is then used for label prediction of test set mutants.}
\label{fig_implementation}
\vspace{-4mm}
\end{figure*}

\section{Approach - \our}
\label{sec_approach}

The main objective of \our is to predict whether a mutant is likely to be vulnerability-mimicking. In order for our approach to be lightweight in terms of engineering and computational effort, we want \our to be able to (a) learn relevant features of \vmm without requiring manual feature definition, and (b) to do so without costly dynamic analysis of mutant executions. To achieve this, we divide our task into two parts: learning a representation of mutants using code embedding technique, and learning to predict based on such embeddings whether or not the represented mutants are \vmm. 

\subsection{Overview of \our}

Figure~\ref{fig_implementation} shows an overview of \our. We divide our approach into three steps that we detail later in this section:
\begin{enumerate}
\item \emph{Building a token representation}: \our pre-processes the original code in order to remove irrelevant information and to produce abstracted code, which is then tokenized to form a sequence of tokens. Each mutant is eventually transformed into its corresponding token representation and undergoes the next step.  
\item \emph{Representation learning}: We train an encoder-decoder model to generate an embedding, a.k.a. vector representation of the mutant. This step is where \our automatically learns the relevant features of mutants without requiring an explicit definition of these features. 
\item \emph{Classification:} \our trains a classification model to classify the mutants (based on their embeddings) as \vmm or not. The true labels used for training the model are obtained by i) replacing the fixed code file with a mutated code file in the project, ii) executing the test suite, iii) checking whether or not the tests failed, and iv) if yes, then matching whether the failed tests are the same as the vulnerability's failed tests. 
\end{enumerate}

It is interesting to note that the mutant representation learned by \our does not depend on a particular vulnerability. \our rather aims to learn properties of the mutants (and their surrounding context) that are generally vulnerability mimicking. This is in line with the recent work on contextual mutant selection \cite{JustKA17, 9677967, ChekamPBTS20} that aims at selecting high-utility mutants for mutation testing. This characteristic makes \our applicable to pieces of code that it has not seen during training. Our results also confirm the capability of \our to be effective on projects not seen during training. Certainly, to make our classifier effective in practice, the selection of the mutant generation technique is important. 
We use \mubert since it produces a sufficiently large set of useful mutants by masking and replacing tokens of the class under analysis. Also, since it employs a pre-trained language model, it proposes code (mutants) similar to the one written by programmers.

\subsection{Token Representation}

A major challenge in learning from raw source code is the huge vocabulary created by the abundance of identifiers and literals used in the code \cite{DBLP:journals/tosem/TufanoWBPWP19, DBLP:conf/icse/TufanoPWBP19, DBLP:conf/iclr/AlonBLY19, DBLP:journals/ese/GargDJCPT22, DBLP:journals/corr/abs-2301-12284}. In our case, this large vocabulary may hinder \our's ability to learn relevant features of \vmm. Thus, we first abstract original (non-mutated) source code by replacing user-defined entities (function names, variable names, and string literals) with generic identifiers that can be reused across the source code file. During this step, we also remove code comments. This pre-processing yields an abstracted version of the original source code, as the abstracted code snippet in Figure~\ref{fig_implementation}.

To perform the abstraction, we use the publicly available tool \emph{src2abs}~\cite{DBLP:conf/icse/TufanoPWBP19}. This tool first discerns the type of each identifier and literal in the source code. Then, it replaces each identifier and literal in the stream of tokens with a unique ID representing the type and role of the identifier/literal in the code. Each ID \textless\texttt{TYPE}\textgreater\texttt{$\_\#$} is formed by a prefix, (i.e., \textless\texttt{TYPE}\textgreater\texttt{$\_$} ) which represents the type and role of the identifier/literal, and a numerical ID, (i.e.,~\texttt{$\#$}) which is assigned sequentially while traversing through the code. These IDs are reused when the same identifier/literal appears again in the stream of tokens. Although we use src2abs, any utility that identifies user-defined entities and replaces such with reusable identifiers can be used as an alternative.

Next, to represent a mutant, we annotate the abstracted code with a mutation annotation on the statement next to the operand/operator that has been mutated. These annotations indicate the applied mutation operation, e.g., \emph{BinaryOperatorMutator} represents mutation on the binary operator ``$>=$", as shown in figure~\ref{fig_implementation}. We repeat the process for every mutant.

Finally, we flatten every mutant (by removing newline, extra white space, and tab characters) to create a single-space-separated sequence of tokens. Using these sequences, we intend to capture as much code as possible around the mutant without incurring an exponential increase in training time~\cite{DBLP:conf/icsm/TufanoWBPWP19, DBLP:conf/icse/TufanoPWBP19, 9677967, DBLP:journals/ese/GargDJCPT22, DBLP:journals/corr/abs-2301-12284}. We found a sequence length of 150 tokens to be a good fit for our task as it does not exceed 18 hours of training time (wall clock) on a Tesla V100 GPU.

\subsection{Embedding Learning with Encoder-Decoder}

Our next step is to learn the embedding, a.k.a. vector representation (that is later used to train a classification model) from mutants' token representation. We develop an encoder-decoder model, a neural architecture commonly used in representation learning task~\cite{DBLP:conf/emnlp/KalchbrennerB13}. The key principle of our encoder-decoder architecture is that the encoder transforms the token representation into an embedding and the decoder attempts to retrieve the original token representation from the encoded embedding. The learning objective is then to minimize the binary cross-entropy between the original token representation and the decoded one. Once the model training has converged, we can compute the embedding from any other mutant's token representation by feeding the latter into the encoder and retrieving the output embedding.

We use a bi-directional Recurrent Neural Network (RNNs) \cite{DBLP:journals/corr/BritzGLL17} to develop our encoder-decoder, as previous works on code learning have demonstrated the effectiveness of these models to learn useful representations from code sequences~\cite{DBLP:journals/corr/BahdanauCB14, 9677967, DBLP:journals/ese/GargDJCPT22, DBLP:conf/nips/SutskeverVL14, DBLP:journals/corr/abs-2301-12284}. We build \our on top of KerasNLP~\cite{kerasnlp2022} which is a natural language processing library providing a general purpose \emph{Transformer Encoder-Decoder} architecture following the work of Vaswani et. al~\cite{DBLP:conf/nips/VaswaniSPUJGKP17} which has shown to perform good both in software engineering and other learning tasks~\cite{DBLP:journals/jaiscr/ShewalkarNL19, DBLP:journals/ese/GargDJCPT22, DBLP:journals/corr/abs-2301-12284}.

To determine an appropriate number of training epochs for model convergence, we conducted a preliminary study involving a small validation set (independent of both the training and test sets used in our evaluation) where we monitor the model's performance in replicating (as output) the same mutant sequence provided as input. We pursue training the model till the training performance on the validation set does not improve anymore. We found 10 epochs for the sequences up to a length of 150 tokens to be a good default for our validation sets.

\subsection{Classifying Vulnerability-mimicking mutants}

Next, we train a classification model to predict whether a mutant, which is represented by the embedding produced by the Encoder, is likely to be \vmm. The learning objective here is to maximize the classification performance, which we mainly measure with Matthews Correlation Coefficient (MCC), Precision, and Recall, as detailed in section \ref{sec_performance_metrics}. To obtain our true classification labels, we replace the fixed code file with a mutated code file in the project, execute the test suite, and check whether or not the tests failed. If the tests fail, we match if the failed tests are the same as the vulnerability’s failed tests to determine whether or not the mutant is a vulnerability-mimicking mutant. For developing the classification model, we rely on \emph{random forests}~\cite{DBLP:journals/ml/Breiman01} because these are lightweight to train and have shown to be effective in solving various software engineering tasks~\cite{Jimenez2019realworld, Pinto2020vocabulary}. We used standard parameters for random forests, viz. we set the number of trees to 100, use Gini impurity for splitting, and set the number of features (i.e. embedding logits) to consider at each split to the square root of the total number of features.

Once the model training has converged, we use the random forest to predict whether a mutant (in the testing set) is likely to be \vmm. We make the mutant go through the preprocessing pipeline to obtain its abstract token representation, then feed this representation into the trained encoder-decoder model to retrieve its embeddings, and input this embedding into the classifier to obtain the predicted label (vulnerability-mimicking or not).


\section{Research Questions}
\label{sec_research_questions}
We start our analysis by investigating how many vulnerabilities in our dataset can be behaviourally mimicked by one or more mutants, i.e., how many mutants fail the same PoVs (tests that were failed by the respective vulnerabilities). Therefore we ask:
\begin{enumerate}
\item [\textbf{RQ1}] \emph{Empirical observation I:} How many vulnerabilities can be mimicked by the mutants? 
\end{enumerate}
For this task, we rely on \vulj dataset~\cite{DBLP:conf/icse/MolinadA22} (section~\ref{sec_vulj}) for obtaining vulnerable projects with vulnerabilities, corresponding fixes, and PoV tests, and on \mubert~\cite{DBLP:conf/icst/DegiovanniP22} (section~\ref{sec_mubert}) for generating mutants. In the \vulj dataset, the fixes (for the vulnerabilities) passed the corresponding project's test suite (containing the PoV tests) in 45 cases for which we mention the details in Table~\ref{tab_data}. \mubert produces mutants of the fixed code, which are checked for mimicking the corresponding vulnerability by replacing the fixed code file with the mutant and executing the test suite. Apart from checking how many vulnerabilities can be mimicked by the mutants, we also analyze how semantically similar the generated mutants are with the vulnerabilities. We measure the semantic similarity of a mutant with the vulnerability by calculating the Ochiai coefficient~\cite{Ochiai} as explained in the following section~\ref{sec_ochiai}. Hence, we ask:
\begin{enumerate}
\item [\textbf{RQ2}] \emph{Empirical observation II:} How similar are the generated mutants with vulnerabilities?
\end{enumerate}
Next, we analyze if the features of \vmm can be automatically learned by machine learning models to statically predict these without the need of investing effort in defining such features. We do so by training models as explained in section~\ref{sec_approach} and check the performance of \our in predicting \vmm. Hence, we ask:
\begin{enumerate}
\item [\textbf{RQ3}] \emph{Prediction Performance:} How effective is \our in automatically defining and learning the features associated with \vmm?
\end{enumerate}


\section{Experimental Setup}
\label{sec_experimental_setup}

\subsection{Semantic similarity}
\label{sec_ochiai}
Mutation seeds artificial faults, a.k.a. mutants, by performing slight syntactic modifications to the program under analysis. For instance, in Figure~\ref{fig_implementation}, the expression $\texttt{x >= y}$ can be mutated to $\texttt{x < y}$. 
Semantic similarity is usually used to evaluate fault seeding~\cite{jia2009, JustJIEHF14, PapadakisSYB18}, i.e. how similar is a mutant (seeded artificial fault) to the desired (real) fault. In the case of this study, the desired fault is the corresponding vulnerability. 

To compute the semantic similarity we resort to dynamic test executions. We use a similarity coefficient, i.e., Ochiai coefficient~\cite{Ochiai}, to compute the similarity of the passing and failing test cases. This is a common practice in many different lines of work, such as mutation testing \cite{jia2009, PapadakisSYB18}, program repair \cite{GouesNFW12}, and code analysis \cite{GoldBHIKY17} studies. Since semantic similarity compares the behavior between two program versions using a reference test suite, the Ochiai coefficient approximates program semantics using passing and failing test cases. 

The Ochiai coefficient represents the ratio between the set of tests that fail in both versions over the total number of tests that fail in the sum of the two. For instance, let $P_1$, $P_2$, $fTS_1$ and $fTS_2$ be two programs and their respective set of failing tests, then the Ochiai coefficient between programs $P_1$ and $P_2$ is computed as:
\begin{align*}
Ochiai(P_1, P_2) = \frac{|fTS_1 \cap fTS_2|}{\sqrt{|fTS_1| \times |fTS_2|}}
\end{align*}

The Ochiai coefficient ranges from 0 to 1, with 0 in case of none of the failed tests is the same between both versions of the programs, (i.e., a mutant and the vulnerability that it is trying to mimic), and 1 in case of all the failed tests match between both versions. Intuitively, a mutant $M$ \emph{mimics} vulnerability $V$, if and only if its semantic similarity is equal to $1$, i.e., $Ochiai(V, M) = 1$.
The mutants shown in Figures \ref{fig_example_1} and \ref{fig_example_2} have an Ochiai coefficient equal to $1$ with their corresponding vulnerability. 

\subsection{Prediction Performance Metrics}
\label{sec_performance_metrics}
\vmm prediction modeling is a binary classification problem, thus it can result in four types of outputs: Given a mutant is vulnerability-mimicking if it is predicted as vulnerability-mimicking, then it is a true positive (TP); otherwise, it is a false negative (FN). Vice-versa, if a mutant does not mimic the vulnerability and, if it is predicted as vulnerability-mimicking then it is a false positive (FP); otherwise, it is a true negative (TN). From these, we can compute the traditional evaluation metrics such as \emph{Precision} and \emph{Recall}, which quantitatively evaluate the prediction accuracy of prediction models.
\begin{align*}
\emph{Precision} = \frac{TP}{TP + FP}  \hspace{1em} \emph{Recall} = \frac{TP}{TP + FN} 
\end{align*}
Intuitively, \emph{Precision} indicates the ratio of correctly predicted positives over all the considered positives. \emph{Recall} indicates the ratio of correctly predicted positives over all the actual positives. 
Yet, these metrics do not take into account the true negatives and can be misleading, especially in the case of imbalanced data. Hence, we complement these with \emph{Matthews Correlation Coefficient (MCC)}, a reliable metric of the quality of prediction models~\cite{DBLP:conf/ease/YaoS20, DBLP:journals/ese/GargDJCPT22}. It is regarded as a balanced measure that can be used even when the classes are of very different sizes~\cite{DBLP:journals/tse/ShepperdBH14}, e.g. \emph{3.9\%} \vmm in total, for 45 vulnerabilities in our dataset (as shown in Table~\ref{tab_distribution}). \emph{MCC} is calculated as: \[\emph{MCC} = \frac{TP \times TN - FP \times FN}{\sqrt{(TP + FP)(TP + FN)(TN + FP)(TN + FN)}}\] 
\emph{MCC} returns a coefficient between 1 and -1. An MCC value of 1 indicates a perfect prediction, while a value of -1 indicates a perfect inverse prediction, i.e., a total disagreement between prediction and reality. MCC value of 0 indicates that the prediction performance is equivalent to random guessing.

\subsection{Experimental Procedure}
\label{subsec:experimentalsettings}
To answer our RQs, we first execute the test suite for every mutant produced by \mubert and analyze which mutants fail the same tests that were failed by the vulnerability to determine \vmm. In total, \mubert produces \emph{16,409} mutants for the fixed versions of the 45 projects (for which the 45 corresponding vulnerabilities are mentioned in Table~\ref{tab_data}). We repeated the test suite execution process for every project to label \vmm that mimic the corresponding vulnerability. 

Once the labeling is complete, to answer \emph{RQ1}, we perform an exact match of the mutant's failed tests with the vulnerability's failed tests to determine how many vulnerabilities are mimicked by the generated mutants. 
To answer \emph{RQ2}, we rely on the Ochiai similarity coefficient (elaborated in Section \ref{sec_ochiai}) to measure how similar the generated mutants are with the vulnerabilities. We calculate the Ochiai coefficient to compute the similarity of the passing and the failing test cases of every vulnerability with all the corresponding project's mutants. To answer \emph{RQ3}, we train models on \vmm and perform k-fold cross-validation (where k = 5) at project level (our dataset has only 1 vulnerability per project) where each fold contains 9 projects. So, we train on mutants of 36 projects (4 training folds) and test on mutants of the remaining 9 projects (1 test fold). Once we get the predictions for all 45 subjects, we compute the Prediction Performance Metrics (elaborated in Section \ref{sec_performance_metrics}) for \our in order to show its learning ability. 


\section{Experimental Results}
\label{sec_experimental_results}

\subsection{Empirical observation I (RQ1)}
\mubert generates \emph{16,409} mutants in total, for all projects in our dataset. Out of \emph{16,409} mutants, \emph{646} mutants are  \vmm mimicking \emph{25} out of \emph{45} vulnerabilities, i.e., at least one or more mutants behave the same as 25 vulnerabilities. Overall, \emph{3.9\%} of the generated mutants mimicked \emph{55.6\%} of the vulnerabilities in our dataset. 
Table~\ref{tab_distribution} shows the project-wise distribution of \vmm  including the total number of mutants generated and the number (and percentage) of mutants that mimic the vulnerabilities. 
These results are encouraging and evidence the potential value of using \vmm as test requirements in practice for security-conscious testing, leading to test suites that can tackle similar mimicked vulnerabilities.  

\begin{tcolorbox}[standard jigsaw, opacityback=0]
\vspace{-1.5mm}
Answer to RQ1: \mubert-generated 646 out of 16,409 mutants mimicked 25 out of 45 vulnerabilities, i.e., \emph{3.9\%} of the generated mutants mimicked \emph{55.6\%} of the vulnerabilities. 
This evidence that pre-trained language models can produce test requirements (mutants) that behave the same as vulnerabilities, making security-conscious mutation testing feasible.
\vspace{-1.5mm}
\end{tcolorbox}

\begin{table}[t!]
\caption{RQ1: The table records the \vmm distribution details that include the total number of generate mutants across all the projects with vulnerabilities, and the number and percentage of \vmm among them. Overall, 3.9\% of the generated mutants mimic 55.6\% of the vulnerabilities.}
\begin{center}
\begin{tabular}{l|r|R{1cm}|r}
\toprule
\multicolumn{1}{c|}{\textbf{CVE}} & \multicolumn{1}{c|}{\textbf{$\#$ Total}} & \multicolumn{2}{c}{\textbf{Vulnerability-mimicking}} \\
\multicolumn{1}{c|}{\textbf{(Vulnerability)}} & \multicolumn{1}{c|}{\textbf{mutants}} & \multicolumn{2}{c}{\textbf{mutants}} \\
 &  & \multicolumn{1}{c|}{\textbf{($\#$)}} & \multicolumn{1}{c}{\textbf{($\%$})} \\ \midrule
CVE-2017-18349 & 286 & 0 & 0\% \\
CVE-2013-2186 & 191 & 0 & 0\% \\
CVE-2014-0050 & 456 & 0 & 0\% \\
CVE-2018-17201 & 375 & 8 & 2.13\% \\
CVE-2015-5253 & 257 & 0 & 0\% \\
HTTPCLIENT-1803 & 553 & 5 & 0.9\% \\
PDFBOX-3341 & 2169 & 308 & 14.2\% \\
CVE-2017-5662 & 511 & 86 & 16.83\% \\
CVE-2018-11797 & 266 & 1 & 0.38\% \\
CVE-2016-6802 & 338 & 16 & 4.73\% \\
CVE-2016-6798 & 441 & 19 & 4.31\% \\
CVE-2017-15717 & 437 & 77 & 17.62\% \\
CVE-2016-4465 & 48 & 0 & 0\% \\
CVE-2014-0116 & 167 & 0 & 0\% \\
CVE-2016-8738 & 50 & 0 & 0\% \\
CVE-2016-4436 & 74 & 0 & 0\% \\
CVE-2016-2162 & 169 & 1 & 0.59\% \\
CVE-2018-8017 & 738 & 17 & 2.3\% \\
CVE-2014-4172 & 212 & 12 & 5.66\% \\
CVE-2019-3775 & 9 & 0 & 0\% \\
CVE-2018-1002200 & 177 & 0 & 0\% \\
CVE-2017-1000487 & 586 & 0 & 0\% \\
CVE-2018-20227 & 18 & 3 & 16.67\% \\
CVE-2013-5960 & 112 & 1 & 0.89\% \\
CVE-2018-1000854 & 9 & 2 & 22.22\% \\
CVE-2016-3720 & 387 & 0 & 0\% \\
CVE-2016-7051 & 387 & 0 & 0\% \\
CVE-2018-1000531 & 158 & 2 & 1.27\% \\
CVE-2018-1000125 & 155 & 14 & 9.03\% \\
APACHE-COMMONS-001 & 144 & 1 & 0.69\% \\
CVE-2013-4378 & 189 & 0 & 0\% \\
CVE-2018-1000865 & 432 & 2 & 0.46\% \\
CVE-2018-1000089 & 205 & 7 & 3.41\% \\
CVE-2015-6748 & 989 & 0 & 0\% \\
CVE-2016-10006 & 356 & 1 & 0.28\% \\
CVE-2018-1000615 & 67 & 38 & 56.72\% \\
CVE-2017-8046 & 12 & 0 & 0\% \\
CVE-2018-11771 & 1754 & 12 & 0.68\% \\
CVE-2018-15756 & 274 & 0 & 0\% \\
CVE-2018-1000850 & 307 & 2 & 0.65\% \\
CVE-2017-1000207 & 29 & 0 & 0\% \\
CVE-2019-10173 & 1658 & 10 & 0.6\% \\
CVE-2019-12402 & 246 & 1 & 0.41\% \\
CVE-2020-1953 & 11 & 0 & 0\% \\
\bottomrule
\end{tabular}
\label{tab_distribution}
\end{center}
\end{table}

\begin{figure*}[t!]
\centerline{\includegraphics[width=1\textwidth]{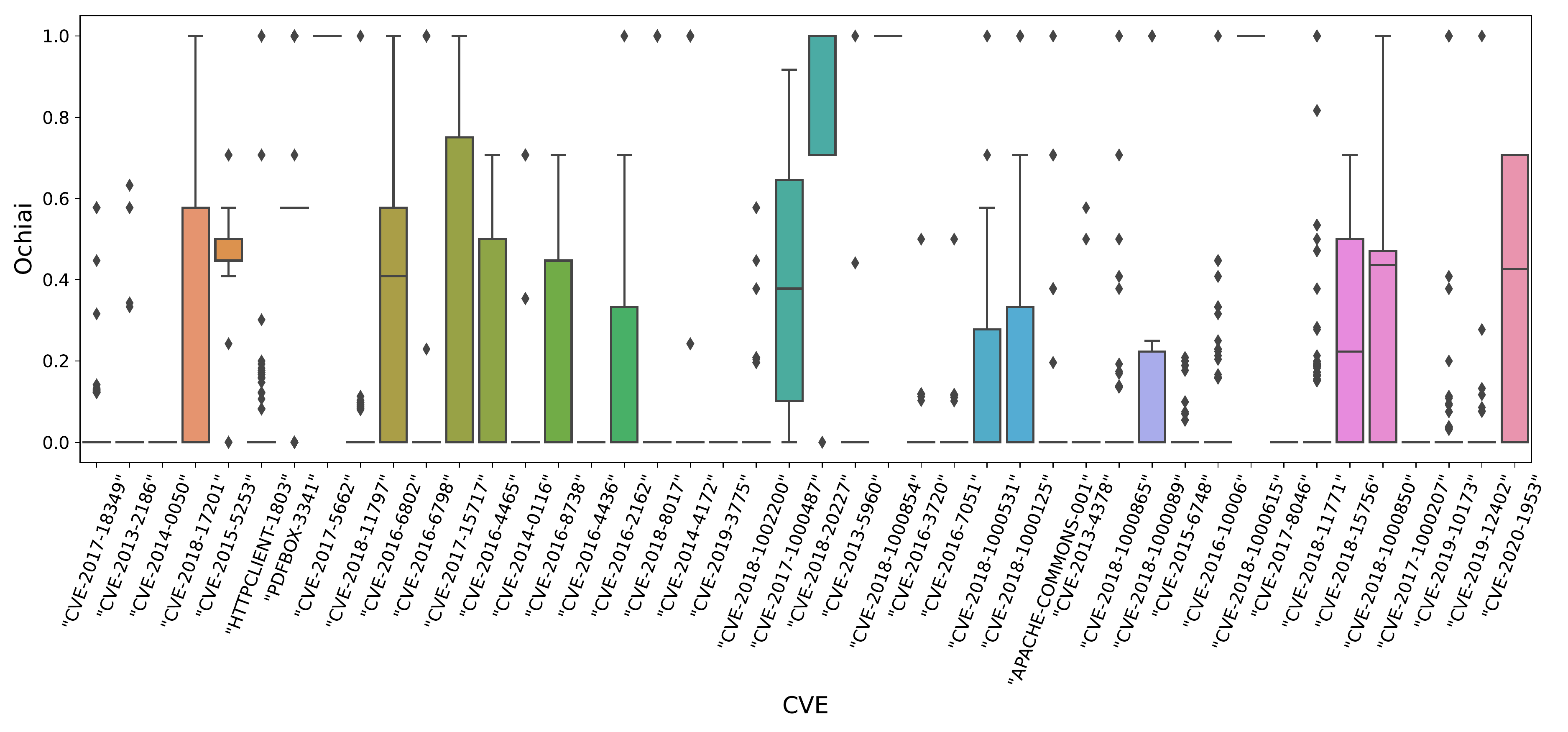}}
\caption{RQ2: Distribution of the mutant-vulnerability similarity in terms of Ochiai similarity coefficient across all the vulnerabilities when compared for similarity with the generated mutants. Overall, 16.6\% of the mutants fail one or more tests that were failed by 88.9\% of the respective vulnerabilities.}
\label{fig_ochiai}
\end{figure*}

\subsection{Empirical observation II (RQ2)}
In addition to 646 mutants mimicking 25 vulnerabilities, i.e., 646 mutants failed by exactly the same tests as the respective 25 vulnerabilities of the corresponding projects, \emph{2,720} mutants achieved the Ochiai similarity coefficient greater than 0 with 40 vulnerabilities of the corresponding projects. This shows that 2,720 mutants, i.e., \emph{16.6\%} of the mutants fail one or more tests (of the corresponding projects) that were failed by 40 respective vulnerabilities, i.e., \emph{88.9\%} of the vulnerabilities in our dataset. Figure~\ref{fig_ochiai} provides an overview of the mutant-vulnerability similarity in terms of Ochiai similarity coefficient distribution across all the vulnerabilities in our dataset when compared for similarity with the generated mutants. 

Despite not behaving exactly the same as the vulnerability, there are many mutants that share some vulnerable behaviors which can help testers to identify the cause of the vulnerability. 
Moreover, vulnerability-similar mutants can help to design more thorough and complete suites to tackle vulnerabilities. 

\begin{tcolorbox}[standard jigsaw, opacityback=0]
Answer to RQ2: \mubert-generated 2,720 out of 16,409 mutants achieved an Ochiai similarity coefficient greater than 0 with 40 out of 45 vulnerabilities, i.e., \emph{16.6\%} of the generated mutants fail one or more tests (of the corresponding projects) that were failed by 88.9\% of the respective vulnerabilities.
\end{tcolorbox}

\subsection{Prediction Performance (RQ3)}
Despite the class imbalance, \our effectively predicts \vmm with a prediction performance of 0.63 MCC, 0.80 Precision, and 0.51 Recall outperforming a random selection of \vmm (i.e., MCC equals 0). These scores indicate that the features of \vmm can be automatically learned by machine learning models to statically predict these without the need of investing effort in defining such features. 
Indeed, any improvement in the mutation testing tools or the pre-trained language models that allow producing better \vmm, can leverage \our to select a more complete set of security-related test requirements. 

\begin{tcolorbox}[standard jigsaw, opacityback=0]
Answer to RQ3: \our achieves a prediction performance of 0.63 MCC, 0.80 Precision, and 0.51 Recall in predicting \vmm. This indicates that the features of \vmm can be automatically learned by machine learning models to statically prioritize these prior to any analysis or execution. 
\end{tcolorbox}

\section{Threats to Validity}
\label{sec_threats_to_validity}
\textit{External Validity}: Threats may relate to the vulnerabilities we considered in our study. Although our evaluation expands to vulnerabilities of severity ranging from high to low, spanning from single method fix to multiple methods modified during the fix (as shown in Table~\ref{tab_data}, the results may not generalize to other vulnerabilities. We consider this threat of low importance since we verify all the vulnerabilities and also their fixes by executing tests provided in the Vul4J dataset~\cite{DBLP:conf/msr/BuiSF22}. Moreover, our predictions are based on the local mutant context, which has been shown to be a determinant of mutants' utility \cite{JustKA17, 9677967}. Other threats may relate to the mutant generation tool, i.e., \mubert that we used. This choice was made since \mubert relies on CodeBERT to produce  mutations that look natural and are effective for mutation tesing. 
We consider this threat of low importance since one can use a better mutant generation tool that can produce more \vmm, which will help \our in achieving better prediction performance. Nevertheless, in case other techniques produce different predictions, one could re-train, tune and use \our for the specific method of interest, as we did here with \mubert mutants. 

\textit{Internal Validity}: Threats may relate to the restriction that we impose on sequence length, i.e., a maximum of \emph{150} tokens. This was done to enable reasonable model training time, approximately \emph{18} hours to learn mutant embeddings on Tesla V100 gpu. Other threats may be due to the use of \emph{Transformer Encoder-Decoder} following the work of Vaswani et. al~\cite{DBLP:conf/nips/VaswaniSPUJGKP17} for learning mutant embeddings. This choice was made for simplicity to use the related framework out of the box similar to the related studies~\cite{DBLP:journals/jaiscr/ShewalkarNL19, DBLP:journals/ese/GargDJCPT22}. Other internal validity threats could be related to the test suites we used and the mutants considered as vulnerability mimicking. We used well-tested projects provided by the Vul4J dataset~\cite{DBLP:conf/msr/BuiSF22}. To be more accurate, our underlying assumption is that the extensive pool of tests including the Proof-of-Vulnerability (PoV) available in our experiments is a valid approximation of the program's test executions, especially the proof of a vulnerability and its verified fix.

\textit{Construct Validity}: 
Threats may relate to our metric to measure the semantic similarity of a mutant and a vulnerability, i.e., the Ochiai coefficient. We relied on the Ochiai coefficient because it is widely known in the fault-seeding community as a representative metric to capture the semantic similarity between a seeded and real fault. In the context of this study, the seeded fault is a mutant and the real fault is a vulnerability. We consider this treat of low importance as the Ochiai coefficient takes into consideration the failed tests of a mutant and a vulnerability (as explained in section~\ref{sec_ochiai}) representing the observable behavior and serving its purpose for this study.

\section{Related Work}
\label{sec:related-work}
The unlikelihood of standard PIT~\cite{pitest} operators to produce security-aware mutants was observed by Loise et al.~\cite{DBLP:conf/icst/LoiseDPPH17} where the authors designed pattern based operators to target specific vulnerabilities. They relied on static analysis for validation of generated mutants to have similarities with their targeted vulnerabilities. 

Fault modeling related to security policies was explored by Mouehli et al.~\cite{4344128} where they designed new mutation operators corresponding to fault models for access control security policies. Their designed operators targeted modifying user roles and deleting policy rules to modify application context, specifically targeting the implementation of access control policies. 

Mutating high-level security protocol language (HLPSL) models to generate abstract test cases was explored by Dadeau et al.~\cite{DBLP:journals/stvr/DadeauHKMR15} where their proposed mutations targeted to introduce leaks in the security protocols. They relied on the automated validation of Internet security protocols and applications tool set to declare the mutated protocol unsafe and capable of exploiting the security flaws.

Targeting black box testing by mutating web applications' abstract models was explored by Buchler et al.~\cite{DBLP:conf/ssiri/BuchlerOP12} where they produced model mutants by removing authorization checks and introducing noisy (non-sanitized) data. They relied on model-checkers to generate execution traces of their mutated models for the creation of intermediate test cases. Their work was focused on guiding penetration testers to find attacks exploiting implementation-based vulnerabilities (e.g., a missing check in a RBAC system, non-sanitized data leading to XSS attacks).

Similar to Loise et al., Nanavati et al.~\cite{DBLP:conf/icst/NanavatiWHJK15} also show that traditional mutation operators only simulate some simple syntactic errors. Hence, they designed memory mutation operators to target memory faults and control flow deviation. They focused on programs in C language and rely on memory allocation primitives in specific to C. Similarly, Shahriar and Zulkernine~\cite{DBLP:conf/compsac/ShahriarZ08} and Ghosh et al.~\cite{DBLP:conf/sp/GhoshOM98} also defined mutation operators related to the memory faults. Their designed operators also exploited memory manipulation in C programs (such as buffer overflows, uninitialized memory allocations, etc.), which security attacks may exploit. These works also focused on programs in C language.

Unlike the above-mentioned related works, we do not target a specific vulnerability pattern/type. Also, since we rely on a pre-trained language model (employed by \mubert), we do not require to design specific mutation operators to target specific security issues. Additionally, our validation of vulnerability-mimicking mutants is not based on a static analysis, but rather a dynamic proof as our produced/predicted vulnerability-mimicking mutants fails tests that were failed by respective vulnerabilities, a.k.a., Proof-of-vulnerability (PoV).

\section{Conclusion}
\label{sec_conclusion}
In this study, we showed that language model based mutation testing tools can produce \vmm, i.e., mutants that mimic the observable behavior of vulnerabilities. Since these mutants are a few, i.e., 3.9\% of the entire mutant set, there is a need for a static approach to identify such mutants. To achieve this, we presented \our, a method that learns to select \vmm from given mutant's code context. Our experiments show that \our identified \vmm with 0.63 MCC, 0.80 Precision, and 0.51 Recall, which indicates that the features of \vmm can be automatically learned by machine learning models to statically predict these without the need of investing effort in defining such features.

\newpage
\bibliographystyle{plain}
\bibliography{Bibliography}

\end{document}